\documentclass{llncs}

\usepackage{graphics}
\usepackage{float}
\usepackage{rotating}
\usepackage{caption}
\usepackage{wrapfig,courier,flafter}
\usepackage{url}
\usepackage{multirow}
\usepackage{booktabs}
\usepackage{subfig}
\usepackage[bottom]{footmisc}
\usepackage[toc,page]{appendix}
\usepackage{fancyhdr}
\usepackage{hyperref}

\usepackage{array}
\newcolumntype{P}[1]{>{\centering\arraybackslash}p{#1}}
\newcolumntype{M}[1]{>{\centering\arraybackslash}m{#1}}

\usepackage{perpage}
\MakePerPage{footnote}

\include{macro-editing}

\begin{document}

\title{Scoring Root Necrosis in Cassava Using Semantic Segmentation}
\titlerunning{Scoring Root Necrosis in Cassava Using Semantic Segmentation}  

\author{Jeremy Francis Tusubira\inst{1} \and Benjamin Akera\inst{1} \and Solomon Nsumba\inst{1} \and Joyce Nakatumba-Nabende\inst{1} \and Ernest Mwebaze\inst{2}}

\authorrunning{Jeremy Francis Tusubira et al.} 
%
\tocauthor{Jeremy Francis Tusubira, Benjamin Akera, Joyce Nakatumba-Nabende, Solomon Nsumba, Ernest Mwebaze}

\institute{Makerere University, Kampala, Uganda,\\
\email{\{tusubirafrancisjeremy,akeraben,snsumba\}@gmail.com, jnakatumba@cis.mak.ac.ug}
\and 
Google AI, Accra, Ghana,\\
\email{emwebaze@gmail.com} }

\maketitle              

\begin{abstract}
Cassava a major food crop in many parts of Africa, has majorly been affected by Cassava Brown Streak Disease (CBSD). 
The disease affects tuberous roots and presents symptoms that include a yellow/brown, dry, corky necrosis within the starch-bearing tissues.  Cassava breeders currently depend on visual inspection to score necrosis in roots based on a qualitative score which is quite subjective. In this paper we present an approach to automate root necrosis scoring using deep convolutional neural networks with semantic segmentation. 
Our experiments show that the UNet model performs this task with high accuracy achieving a mean Intersection over Union (IoU) of 0.90 on the test set. This method provides a means to use a quantitative measure for necrosis scoring on root cross-sections. This is done by segmentation and classifying the necrotized and non-necrotized pixels of cassava root cross-sections without any additional feature engineering.
\end{abstract}

\section{Introduction}
Cassava (\textit{Manihot esculenta} Crantz) is a major subsistence crop in many parts of the world. However, in recent years cassava production in Uganda has been affected by Cassava Brown Streak Disease (CBSD) \cite{abaca2012progression,alicai2007re}.
CBSD has characteristic symptoms that appear as chlorotic leaf symptoms along the major veins and necrotic lesions on part or all of the starchy root that appear as yellow/brown \cite{alicai2007re,kawuki2019alternative}. CBSD affects the cassava root tubers which are the edible part of the cassava plant and affected, the roots are unsuitable for consumption \cite{hillocks2016disparity}.
However, there exists poor phenotypic associations between the above-ground CBSD symptoms shown in the leaves and the below-ground CBSD symptoms shown in the roots \cite{kawuki2019alternative,kaweesi2014field,valentor2018plot}. Due to this, the assessment of CBSD entirely on leaves is not sufficient enough to assess for CBSD resistance and that is why the CBSD assessment in this paper is on the roots instead of the leaves.

Efforts to breed for CBSD resistance are being put in place in the East African region \cite{ozimati2018training}. Other approaches to deal with CBSD include disease surveillance and deployment of more resistant varieties of cassava
\cite{katono2015influence,ninsiima2018automating}. Currently, the assessment of CBSD severity on cassava root cross-sections is done by visual analysis by agricultural experts. To assess a cassava root tuber for necrosis, the expert uproots a cassava plant and slices off clean discs from its root tuber and assigns a score of necrosis severity to the tuber. The characteristic symptoms in cassava root tubers are used as a measure of either resistance or tolerance to CBSD \cite{abaca2012progression}.

The current scoring approach is subjective, and moreover different authors also assign different scores for the same scale \cite{katono2015influence,ozimati2018training,rwegasira2012response} which are liable to human bias and can sometimes be erroneous. This could have several effects especially for the cassava breeders who require accurate scores when testing for CBSD resistance \cite{ninsiima2018automating,tuhaise2014pixel}. 
 Therefore, there is need for automated techniques for scoring of necrosis on cassava root cross-sections. 
 
Previous work has been proposed where computer vision techniques are used to calculate the percentage of necrosis on a cassava root infected with necrosis \cite{ninsiima2018automating,tuhaise2014pixel}. However, this relied on hand engineered features like pixel intensity along with traditional classification techniques in a controlled setting, to detect necrosis. Generally, automation of necrosis scoring is a difficult problem because the background of the root has to be segmented out first before determining the necrotized pixels in the root. Furthermore, changes in illumination of images can have a significant impact on pixel intensities of an input image. This paper introduces an approach for determining the percentage of  necrosis on a cassava root image using semantic segmentation. Semantic segmentation has been used in several applications such as phenotyping of plant diseases \cite{stewart2019quantitative}, land cover mapping \cite{kampffmeyer2016semantic} and land use classification \cite{yang2018classification} and several applications in the medical field such as segmentation of brain gliomas from MRI images \cite{cui2018automatic}. 
Semantic segmentation is a challenging task in computer vision. A lot of methods have been developed to achieve this task many of which have been built using deep learning models that have shown the most noticeable performance \cite{He2017ICCV}.

The main contribution of this work is the development of a more robust approach to  necrosis scoring. The paper discusses the use of convolutional neural network, UNet model to automatically segment and classify necrotized and non-necrotized pixels of cassava root cross-sections from the input images obtained from the field. The result of which is a percentage estimate of the level of necrosis obtained from the ratio of necrotized pixels to root pixels. Images of cassava root cross-sections are annotated with the help of experts to generate a training dataset for the model. The results show that the UNet model is able to detect necrosis on input images with a high accuracy. 

The rest of the paper is organised as follows: Section \ref{related_work} discusses related work. Section \ref{methods} describes the methodology used in this paper. Section \ref{model} describes the model architecture and evaluation metrics. Section \ref{experiments} provides a discussion on the experiments and the results from the experiments. Section \ref{conclusion} concludes the paper.

\section{Related Work}\label{related_work}
In this section, we discuss related approaches to necrosis detection and applications of semantic segmentation in agriculture. Current approaches to root necrosis scoring provide a qualitative score
on a scale of 1-5 \cite{hillocks2000cassava} representing how necrotized the cassava root cross-section is \cite{kaweesi2014field,kawuki2016eleven}. In this range, different authors assign different scales for necrosis \cite{katono2015influence,ozimati2018training,rwegasira2012response}. For example, authors in \cite{rwegasira2012response} give score 1 if the root is not necrotic, score 2 is 0-5\% of the root is necrotic, score 3 if 5-30\% of the root is necrotic, score 4 if 30-50\% of the root is necrotic and score 5 if $>$50\% of the root is necrotic. While the authors in \cite{katono2015influence} give score 1 if the root is not necrotic, score 2 is $<$5\% of the root is necrotic, score 3 if 5-10\% of the root is necrotic, score 4 if 10-25\% of the root is necrotic and score 5 if $>$25\% of the root is necrotic. Although they refer to the same scale, the expression of necrosis under each scale is different. 

Work has been carried out to automate plant disease diagnosis and symptom measurement using image analysis techniques \cite{ramcharan2017deep,mohanty2016using,mwebaze2016prototypes}. With the successes of deep neural networks in supervised machine learning \cite{deng2013new}, convolutional neural networks have played a very significant role in image classification \cite{krizhevsky2012imagenet,simonyan2014very,szegedy2015going} and object detection \cite{redmon2016you,ren2015faster}. This has inherently driven research towards developing convolutional neural network architectures for Semantic Segmentation \cite{pathak2014fully}. Semantic Segmentation based on convolutional neural networks have outperformed other techniques and have thus been applied in many domains including Biomedical research to for identification of cell nuclei \cite{ronneberger2015u}, self-driving cars \cite{treml2016speeding} and remote sensing \cite{kampffmeyer2016semantic}.

Similarly, in agricultural research, segmentation has been applied in cotton detection \cite{li2016field} and for grape detection and tracking \cite{santos2019grape}. It was also been applied in disease detection. In \cite{singh2018deep}, semantic segmentation has been applied to phenotype plant stress. Quantitative phenotyping of the northern leaf blight virus has been successfully carried out by applying semantic segmentation to images captured by unmanned aerial vehicles \cite{stewart2019quantitative}. A different approach in \cite{wang2019segroot} has proposed a method based on neural networks to segment roots from soil background and using images of soybean roots and the results achieved a dice score of 0.64.

In the area of CBSD scoring, authors in \cite{tuhaise2014pixel} carry out analyses of necrosis in cassava root cross-sections at a pixel level using binary pixel classification with traditional classification algorithms in a controlled setting. The work in \cite{ninsiima2018automating} builds on this previous work where the analysis of necrosis is carried out on cassava root cross-sections that are captured in the field setting. Here, hand-crafted features are extracted from the input image to isolate the root from the background and two methods are applied for necrosis detection: Otsu thresholding \cite{otsu1979threshold} with blob detection and the watershed algorithm \cite{gauch1999image} with selected components based on contours present on the root cross-section.

In this paper, we utilize the ability of convolutional neural networks to automatically extract features from images beyond the pixel intensity values used by thresholding methods. This gives the model the ability to identify multiple expressions of necrosis beyond what traditional thresholding methods can do.

\section{Methodology}\label{methods}
\subsection{Data Collection}
In this experiment, 1,800 images of cassava root cross-sections were collected from the National Crop Resources Research Institute (NaCRRI), that hosts the national cassava breeding program of Uganda. Root necrosis severity was assessed on each root by slicing the root transversely 5-7 times. The cross-section with the highest visual expression of necrosis was scored by human experts on a scale of 1-5 as shown in the Table \ref{tab:valid} representing the expression of CBSD on the root  \cite{ozimati2018training}.
For each root, a picture of the cross-section that exhibited the highest score of necrosis was taken using a smartphone camera with a resolution of 1920x2520 pixels under ordinary field conditions and the corresponding CBSD score was recorded.

\begin{table}[!h]
\centering
 \caption{Scoring of Necrotized Regions on Cassava Root Cross-sections.}
 \label{tab:valid}
 \begin{tabular}{ccc}
   \hline
   Human Expert  & Expected Root  & Expected Machine \\
   Score &  Cover (\%) &  Score (\%)\\
   \hline
   1 & no necrosis & 0-2\\
   2 & $\leq$ 5 is necrotic & $\leq$ 5 \\    
   3 & 6-10 is necrotic & 6-10\\
   4  & 11-25 is necrotic & 11-25\\
    5 & $>$25 is necrotic & $>$25 \\
  \hline
\end{tabular}
\end{table}

\subsection{Data Description \label{sec:datades}}
\begin{figure}[h!]
\begin{center}
\includegraphics[width=7cm]{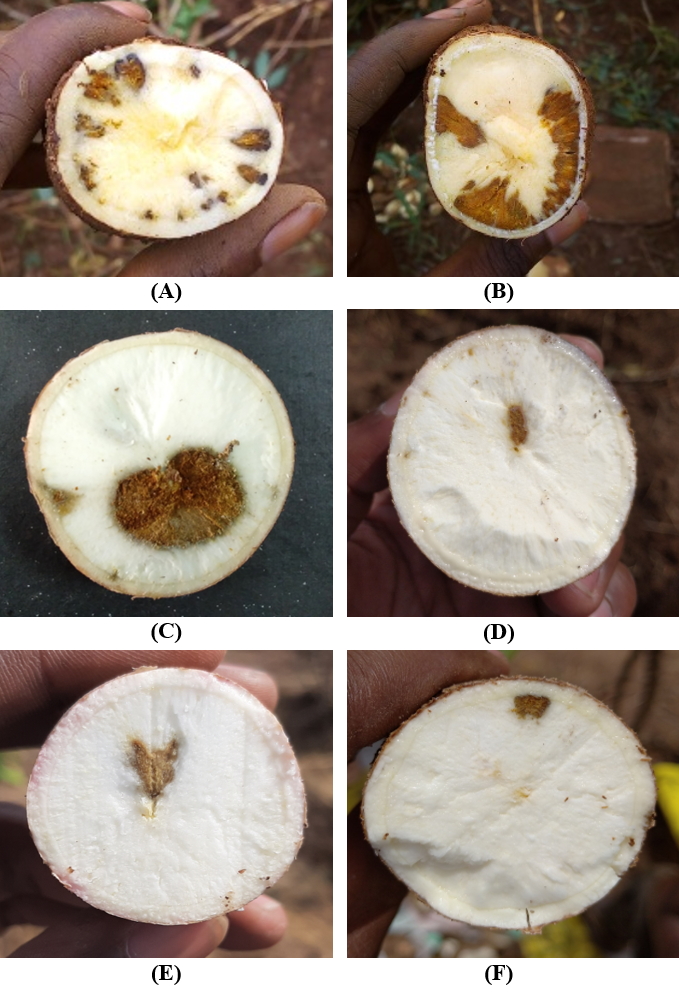}
\end{center}
\caption{Categories of necrosis expressions on cassava roots. \textbf{(A)} many necrosis lesions, \textbf{(B)} few necrosis lesions, \textbf{(C)} large necrosis lesion, \textbf{(D)} small necrosis lesion, \textbf{(E)} necrosis lesion in the center of the root, and \textbf{(F)} necrosis lesion at the edge of the root.}\label{fig:necrosis_category}
\end{figure}

In the dataset, 1,036 images of cassava root cross-sections were necrotic while 764 were clean and did not show any necrosis.
A cross-section of a clean root is white or yellow fleshed  while a necrotic root can expresses white or brown lesions depicting CBSD. 
In this data set, the manifestation of lesions on the necrotic roots is observed to fall under three main categories: (a) number of necrotic lesions - few lesions vs. many lesions, (b) size of the necrotic lesions - small lesions vs. large lesions, (c) the distribution of the necrosis - lesions at the center of the root vs. lesions at the edge of the root. A root can have necrosis presentation that falls under one of more of these categories described as shown in Figure \ref{fig:necrosis_category}. 

\subsection{Data Annotation}
\begin{figure}[ht!]
\centering 
\includegraphics[width=\linewidth]{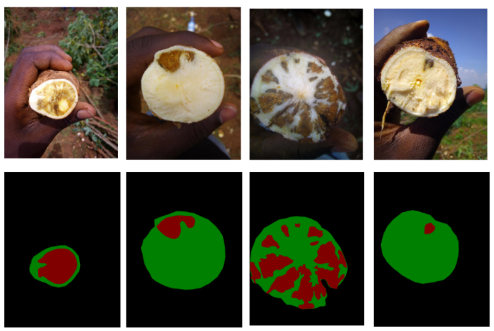} 
\caption{Sample input images of cassava root cross-sections and the actual masks generated from annotation.}  
\label{fig:train} 
\end{figure}

For the segmentation task, the 1,036 images of cassava roots containing necrosis were labelled and split to create the training set with 90\% samples and the validation set with 10\% of the samples. An  additional 128 images were annotated and used as a test set. The ground truth for the segmentation task is a series of pixel-wise masks. The masks were hand-traced by volunteers with the guidance of experienced cassava breeders using the tool \textit{labelMe}\footnote{http://labelme.csail.mit.edu/Release3.0/}. During the annotation, the volunteers drew labelled polygons around two objects of interest: the root area and the necrotic area. The polygons were then converted into masks representing the necrosis lesions, the root and the background. An example of the resulting masks from the annotation are shown in Figure \ref{fig:train}. 
The annotated images were saved in the PASCAL VOC format \cite{everingham2015pascal} which is a standard format for annotating images for the task of object detection and image segmentation. The main resultant files include the original image, generated mask and JSON file containing pixel coordinates.

\section{Model Architecture}\label{model}
\begin{figure}[ht!]
\centering 
\includegraphics[width=\linewidth]{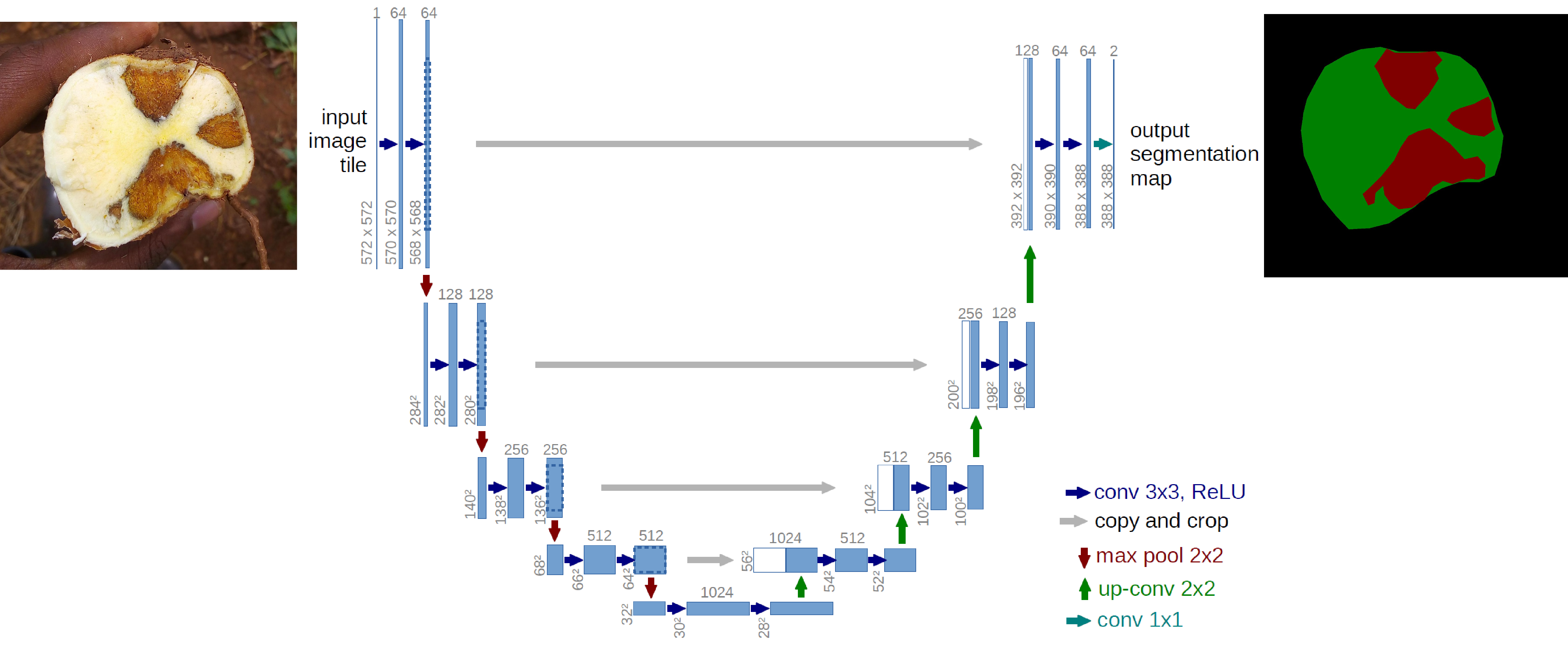} 
\caption{UNet Architecture for semantic segmentation of cassava root cross-sections.}  
\label{fig:unet} 
\end{figure}

For the semantic segmentation task, we used the UNet model \cite{ronneberger2015u}, a commonly used deep-learning architecture for performing image segmentation tasks \cite{zhao2019use}. The architecture of the UNet model is based on an encoder-decoder model with a contracting and expansive arm as shown in Figure \ref{fig:unet}. The UNet architecture has specifically proved to outperform previous methods on the task of segmentation of biomedical images. It has also has the capability of generalizing better even when trained end-to-end on very few images.
 
The model is composed of a linear stack of Convolutional, Batch Normalization and Rectified Linear Units (ReLu) operations followed by a max-pooling operation applied to an image with a 512x512 pixel input dimension. At each pooling layer in the encoder, spatial resolution of the feature map is reduced by 2 and number of feature channels is doubled. We keep track of the outputs of each block as we feed these high-resolution feature maps into the decoder portion.

The decoder portion is comprised of up-sampling convolutional layers together with convolutional and ReLu operations. At each block, a 2x2 up-sampling convolutional layer reduces the number of feature channels by half and the result is concatenated with the corresponding feature map from the encoder. This is followed by convolution and ReLu operations except at the final layer where a 1x1 convolution is used map the final feature map to a certain number of output classes.

For this experiment, an image with an input dimension of 256x256 pixels was used while the rest of the architecture was left unmodified as in \cite{ronneberger2015u}. The model adopts the dice coefficient as shown in equation \ref{dicecoeff} for evaluation. This calculates the ratio of the intersection of the area between the input mask and predicted mask to the sum of the total area of the input mask and the predicted mask. The dice coefficient is scored on a scale of 0-1 where scores closer to 1 imply a greater precision for the predicted mask.  

    \begin{equation}
    \label{dicecoeff}
         \mathrm{DiceCoefficient} = \frac{2(R_{pred} \cap R_{inp})}{R_{pred} + R_{inp} }
    \end{equation}
    
    \begin{equation}
    \label{diceloss}
        \mathrm{DiceLoss} = 1 - \mathrm{DiceCoefficient}
    \end{equation}
    
     \begin{equation}
    \label{iou}
        \mathrm{IOU} = \frac{R_{pred} \cap R_{inp}}{R_{pred} \cup R_{inp} }
    \end{equation}
    
    $R_{pred}$ is the area of the predicted mask and $R_{inp}$ is the area of the input mask.
    
The dice loss function was used to monitor and optimize the parameters of the model during training. The dice loss was defined as the loss function and this can be calculated from the dice coefficient as shown in Equation \ref{diceloss}.
We use the mean Intersection over Union (IoU) as shown in Equation \ref{iou} \cite{milletari2016v} to evaluate how well the model classifies the necrotic and root pixels. The network predictions, which consist of generated masks have the same resolution as the input data, are processed through a SoftMax activation function which predicts which outputs the probability of each pixel belonging to the necrosis, the root or the background class. 

\section{Experiments}\label{experiments}
The problem of calculating the percentage of necrotized roots is addressed as a pixel classification task where we attempt to classify the pixels covered by the necrotized region of the root against the healthy section of the root. Based on this, the severity score on the root is determined by calculating the ratio of the necrotized region to the healthy section.  

\subsection{Training}
For model training, an original image is used as the input feature and the corresponding mask generated by the annotation is the label which is typical of every supervised machine learning task. All the inputs are resized to a dimension of 256x256 pixels which is the input size for the UNet model. The original label masks have a unique pixel value for the 3 regions of interest; root, necrosis and background encoded in an RB color space from 0-128. Before parsing the data for training, the color values of the mask were remapped to a new colorspace representing each of the 3 regions with a unique value in the range 0-2. In order to increase size and variations in the dataset, augmentation was done using three image transformations applied to random selections of input images and their corresponding masks; horizontal flip, rotation and width-height shifts.

The UNet model was trained using Adam \cite{kingma2014adam} optimizer with the initial learning rate set to 3e-4 and weight initialization was implemented using Xavier initialization defined by Glorot and Bengio \cite{glorot2010understanding}. We defined two callbacks; one to save training weights for best performing epoch based on the dice coefficient and another to implement early stopping if the value of the dice coefficient did not change for 20 epochs. The model was trained for a maximum for 100 epochs using Google Colab hosted environment with 12GB of memory powered by an Nvidia K80 Graphics Processing Unit (GPU).

\subsection{Predictions and Inference}
As shown in Figure \ref{fig:unet}, the output of the UNet model is an array with the same dimensions as the input image. When an image is passed into the model for inference, each of the pixels from the input is classified as being one of either two classes; the root or necrosis and everything else is the background. The pixel values of each class from the model prediction are assigned a unique value and the percentage of necrosis on the root is calculated by getting the ratio of pixels representing necrosis to the total number of pixels representing both the root and necrosis as in Equation \ref{necpercentage}.

 \begin{equation}
    \label{necpercentage}
        \mathrm{Necrosispercentage} = {\frac{P_{nec}}{(P_{nec} + P_{root}) }}  \times 100
    \end{equation}
$P_{nec}$ is the number of pixels classified as necrosis and $P_{root}$ is the number of pixels classified as the root.

In order to generate an output image, positions of pixels classified as the root and necrotic areas are extracted from the input image and a black mask with the same dimension as the input image is created. The positions of pixels in the black mask that correspond to those extracted from the input image are modified to reflect the two classes with necrotic pixels set to red and root pixels set to green as shown in Figure \ref{fig:unet_performance}.

\subsection{Performance of the UNet Model}
\begin{figure}[ht!]
\centering 
\includegraphics[width=\linewidth]{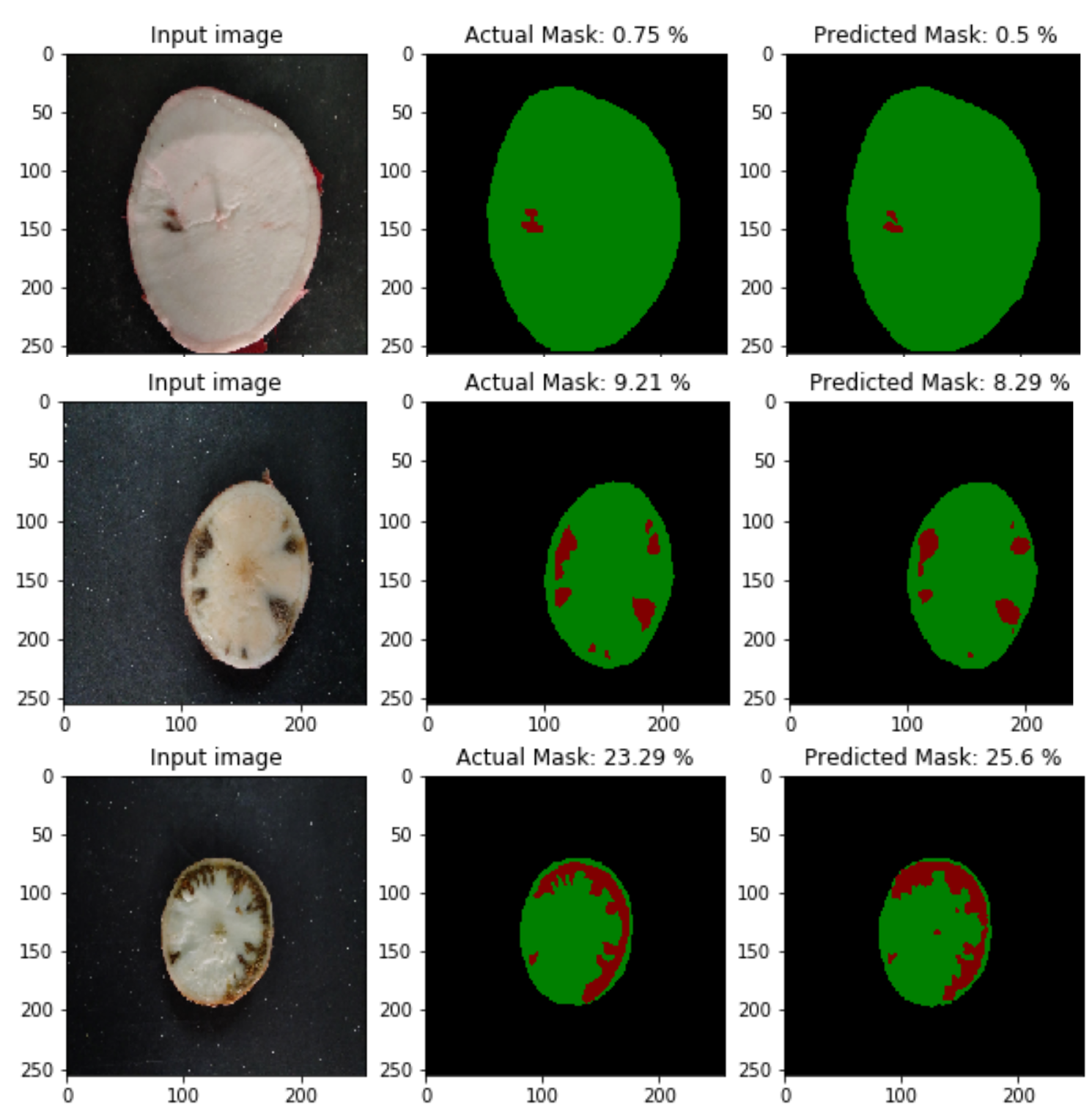} 
\caption{Segmentation results of UNet for three image samples from the test set. The left image shows original RGB image, the middle image shows mask generated by annotation and the right image shows mask generated by prediction of the UNet model.}  
\label{fig:unet_performance} 
\end{figure}

The UNet model showed an impressive performance with dice coefficients of 0.97 and 0.95 for the training and validation sets respectively. This implies that the model was able to correctly learn and identify the necrosis on most examples of the root with minimal mistakes. The model also managed to achieve an IOU of 0.94 and 0.90 for the training and validation sets respectively. Both the dice coefficient and IOU scores show a small difference between the training and validation sets of 0.02 and 0.04 respectively and this implies that the model is robust to effects of overfitting. By visual assessment, Figure \ref{fig:unet_performance} shows that the UNet model is able to classify the necrosis and the root with a high accuracy for the randomly selected sample of 3 images from the test set when the actual mask is compared with the predicted mask. 

\subsection{Comparison with Ground truth}
In this experiment, the performance of the UNet model\footnote{\url{http://18.221.68.135}} is compared to previous work by \cite{ninsiima2018automating} based on how well each of the approaches can estimate the percentage of necrosis on an unseen test set. Using the test set of 128 images, the ground truth percentage of necrosis for each image is generated by computing the pixel based ratio using the $\mathrm{Necrosispercentage}$ shown in Equation \ref{necpercentage}. This is calculated on the masks generated by the ground truth annotations as shown in Figure \ref{fig:train}. 

The test images were then analyzed by the UNet model to generate mask predictions whose pixel ratio is used to calculate the UNet percentage of necrosis. Similarly the test images were analyzed for the Otsu predicted percentage of necrosis by running them through the CBSD analyzer\footnote{\url{http://air.ug/mcrops/?page_id=46}}, a desktop application developed by \cite{ninsiima2018automating} to house the Otsu-threshold model. Using the MSE (Mean Squared Error) and R\textsuperscript{2} as evaluation metrics, the two models were assessed to estimate how much each of the predicted percentages varies from the ground truth percentage and results are presented in Table \ref{tab:otsu_vs_unet}.
 \begin{table}[!h]
\begin{center}
 \caption{MSE, $R\textsuperscript{2}$ and $r$ scores for UNet model and Otsu-threshold model.}
 \label{tab:otsu_vs_unet}
 \begin{tabular}{m{4cm} m{1cm} m{1cm} m{1cm}}
   \hline
    & MSE & $R\textsuperscript{2}$ & $r$\\
   \hline\\
   UNet model  & 72.34 & 0.73 & 0.92\\
    Otsu-threshold model  & 273.96 & -0.03 & 0.44\\
      \hline
\end{tabular}
\end{center}
\end{table}

\begin{figure}[ht!]
\centering 
\includegraphics[width=\linewidth]{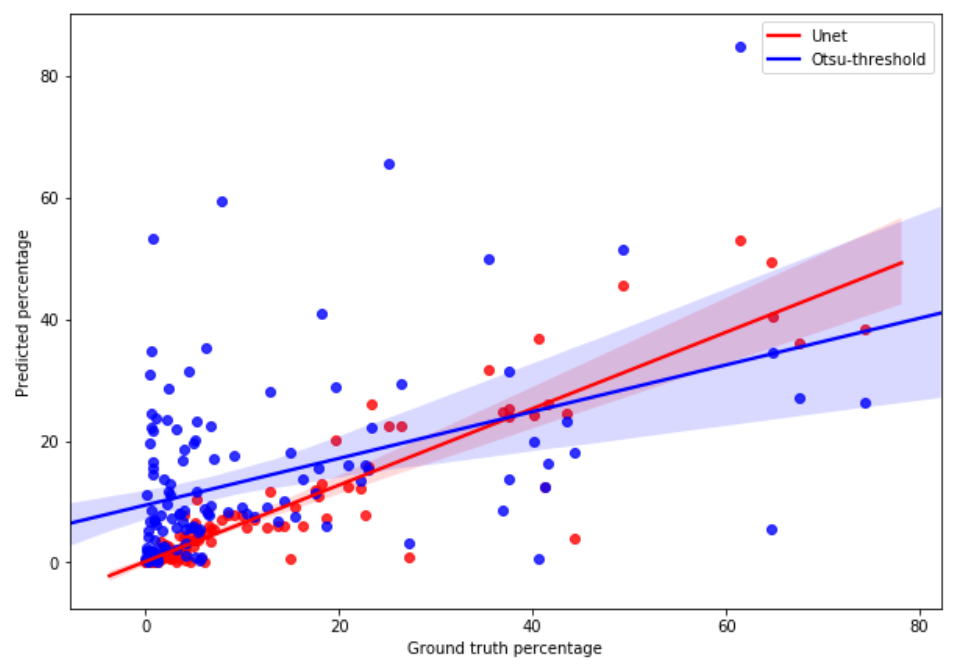} 
\caption{Scatter plot for ground truth percentage against UNet predicted percentage (red) and Otsu-threshold predicted percentage (blue) with corresponding lines of best fit.}  
\label{fig:otsu_unet_regression} 
\end{figure}

The predicted percentage of necrosis for the UNet and Otsu-threshold models were also plotted against the ground truth percentage in a scatter diagram Figure \ref{fig:otsu_unet_regression}. The Pearson correlation coefficient (r) was calculated to determine how close each of the model's predictions were to the ground truth. The correlation coefficients scored by the UNet model and the Otsu-threshold are 0.92 and 0.44 respectively. Visually, Figure \ref{fig:otsu_unet_regression} also shows that the UNet model (red) has a better line of best fit compared to the Otsu-model (blue).

Figure \ref{fig:otsu_unet_regression} also shows a higher correlation between the predicted necrosis percentage and the ground truth for the UNet model compared to the Otsu-threshold model. 

\subsection{Discussion of Results}
In this paper, we have presented results for automatic necrosis scoring of cassava root cross-sections using semantic segmentation. The result show that necrosis scoring is a feasible task that can be accomplished with a significantly high accuracy. When compared with previous work, this paper has presented a richer test set of 128 images compared to the 20 images reported in  \cite{ninsiima2018automating}. Furthermore, the UNet model performs significantly better at the task of necrosis detection when using the annotated test dataset of 128 images as a baseline. 

Unlike the Otsu-threshold model whose performance is greatly influenced by features such as the determined thresholding value and image illumination effects, the UNet model is able to learn and understand a more robust set of features from the input data and this enables it to perform well even in varying image conditions. Based on the results in Table \ref{tab:otsu_vs_unet}, both the $R\textsuperscript{2}$ and $r$ values for the UNet model show that the predictions are closer to the ground truth representation of necrosis. The results from the training experiments show that the UNet model is able to learn and predict necrosis present on cassava root cross-sections with high dice coefficients of 0.96 and 0.94 for the training and validation sets respectively and therefore this is a method that can be employed to automate the current manual approach to scoring necrosis in cassava root tubers.

\section{Conclusion and Future Work}\label{conclusion}
In this paper, we present an approach to detect necrosis in images of cassava root cross-sections using semantic segmentation. This was done with the UNet model which was trained on a dataset of 932 images and evaluated on 104 images. The experimental results show that the model is capable of identifying both the root and necrosis with a high level of accuracy as indicated by the IOU scores of 0.94 and 0.90 for the training and validation sets respectively.

Furthermore, when evaluated on a never seen set of 128 images, the UNet model performs significantly better than the previous method suggested by \cite{ninsiima2018automating} at predicting the actual percentage of necrosis over the entire test set. This shows that the UNet model learns more complex features that are able to generalize to greater variations in root cross-section images compared to methods that are using hand crafted features such as pixel value intensity.

Although the model produces very good results for necrosis detection, we observed some limitations with the analysis of results for the test set. The percentage-based comparison to the ground truth for necrosis detection is not a sufficient measure for how well the masks predicted by the models relate to the ground truth masks. As future work, we will use the IOU or dice coefficient metrics to give a more robust evaluation. 

Future work will also involve investigating the performance of the UNet model for the different types of expression of necrosis discussed in Section \ref{sec:datades}. Using the segmentation results, we shall also be able to incorporate measures like size and number of lesions which could provide useful information to cassava breeding experts. We believe that this model can easily be adapted for use in similar problems like necrosis scoring detection in sweet potato root cross-sections with minimal modification.

\subsubsection*{Acknowledgments.} We would like to Dr. Robert Kawuki and the technical team from the National Crop Resources Research Institute (NaCRRI). The work is funded by Cornell University through a sub-award agreement (N0. 84941-11036) between the Makerere Artificial Intelligence Lab and Cornell University.

%
%

\end{document}